\documentstyle[12pt,a4wide]{article}
\input epsf

\newcommand{\news}{\setcounter{equation}{0}}
\newcommand{\be}{\begin{equation}}
\newcommand{\ee}{\end{equation}}
\newcommand{\bea}{\begin{eqnarray}}
\newcommand{\eea}{\end{eqnarray}}
\newcommand{\bean}{\begin{eqnarray*}}

\newcommand{\eean}{\end{eqnarray*}}
\font\upright=cmu10 scaled\magstep1
\font\sans=cmss12
\newcommand{\ssf}{\sans}
\newcommand{\stroke}{\vrule height8pt width0.4pt depth-0.1pt}
\newcommand{\Z}{\hbox{\upright\rlap{\ssf Z}\kern 2.7pt {\ssf Z}}}

\newcommand{\C}{{\rlap{\rlap{C}\kern 3.8pt\stroke}\phantom{C}}}
\newcommand{\R}{\hbox{\upright\rlap{I}\kern 1.7pt R}}
\newcommand{\CP}{\C{\upright\rlap{I}\kern 1.5pt P}}
\newcommand{\PP}{\hbox{\upright\rlap{I}\kern 1.5pt P}}

\newcommand{\identity}{{\upright\rlap{1}\kern 2.0pt 1}}

\begin{document}
\pagestyle{plain}
\title{\vskip -70pt
\begin{flushright}
\end{flushright}\vskip 50pt
{\bf \Large  \bf A SKYRME LATTICE \\ WITH HEXAGONAL SYMMETRY}\vskip 10pt}

\author{Richard A. Battye$^{\ \dagger}$ and Paul M. Sutcliffe$^{\ \ddagger}$\\[10pt]
{\normalsize $\dagger$ {\sl Theoretical Physics Group, Blackett Laboratory, Imperial College,}}\\{\normalsize {\sl Prince Consort Road, London, SW7 2BZ, UK.}}\\{\normalsize {\sl Email : R.Battye@ic.ac.uk}}\\ 
\\{\normalsize $\ddagger$ {\sl Institute of Mathematics, University of Kent at Canterbury,}}\\
{\normalsize {\sl Canterbury, CT2 7NZ, U.K.}}\\
{\normalsize{\sl Email : P.M.Sutcliffe@ukc.ac.uk}}\\}
\date{}
\maketitle
\vskip 25pt
\begin{abstract}
\noindent Recently it has been found that the structure of
 Skyrmions has a close analogy to that of fullerene
shells in carbon chemistry. In this letter we show that this analogy 
continues further, by presenting a Skyrme field that describes a lattice
of Skyrmions with hexagonal symmetry.
This configuration, a novel `domain wall' in the Skyrme model, has low 
energy per baryon (about 6\%  above the Faddeev-Bogomolny bound) and in 
many ways is analogous to graphite. By comparison to 
the energy per baryon of other known Skyrmions and also the Skyrme crystal, 
we discuss the possibility of finding Skyrmion shells of higher charge.
\end{abstract}
\newpage

\section{Introduction}
\news
Our recent work \cite{BSb}, involving numerical simulations
of the full nonlinear field equations of the Skyrme model, has revealed that
the structure of minimal energy multi-Skyrmions has a rich and fascinating
complexity, where surfaces of constant baryon density are given by trivalent polyhedra with holes at the centre of each face. These structures are very reminiscent of those appearing in fullerene chemistry to describe closed shells of carbon atoms \cite{Kr,FM}, leading us to draw an analogy between Skyrmions and carbon chemistry.

More explicitly, the empirical Geometric Energy Minimization rule of ref.\cite{BSb} states that, the baryon density isosurface of a charge $B$ Skyrmion is an almost spherical trivalent polyhedron with $4(B-2)$ vertices, $2(B-1)$ faces and $6(B-2)$ edges. For $B\ge 7$
the polyhedron comprises twelve pentagons and $2(B-7)$ hexagons, which is precisely the structure of a fullerene corresponding to a closed shell containing $4(B-2)$ carbon atoms. In particular, the $B=7,8$ and 9 minimum energy configurations were seen to have the same structure as shell-like forms of ${\rm C}_{20}$, ${\rm C}_{24}$ and ${\rm C}_{28}$ respectively \cite{BSb,FM}.
Motivated by this numerical work, a new approach to constructing Skyrme fields \cite{HMS}, based upon rational maps between
Riemann spheres was developed. This has allowed a good mathematical understanding of some aspects of these Skyrmions
and has also produced a Skyrme field of charge seventeen which has the structure
of the most famous of the fullerenes, the ${\rm C}_{60}$ Buckminsterfullerene. It too
appears to be the minimal energy Skyrmion of this charge.

In considering very large fullerenes, where hexagons are dominant, the twelve pentagons
may be viewed as defects, inserted into a flat hexagonal structure, in order
to generate the required curvature necessary to close the shell. Energetically the optimum
structure is an infinite hexagonal lattice, that is, a sheet of graphite; the most stable form of elemental carbon from the thermodynamic point of view. The reason that closed shells are preferred
for a finite number of carbon atoms is that the penalty for introducing the pentagonal defects
is not as severe as that incurred by having dangling bonds at the edges of a truncated graphite
sheet. A prediction of the fullerene approach to Skyrmions is therefore that a Skyrme field
should exist which represents a hexagonal lattice, that is, the analogue of graphite. Furthermore,
although this configuration would have infinite energy, since it has infinite extent in two
directions, its energy per baryon should be lower than that of any of the known finite
energy Skyrmions. In the next section we shall verify this prediction and thus provide another
piece of evidence in support of the similarities between fullerenes and Skyrmions.

For completeness, we include expressions for  the Lagrangian of the Skyrme
 model, which in terms
of the $su(2)$-valued right currents $R_\mu=(\partial_\mu U)U^\dagger$ is
\be
{\cal L} =-\displaystyle{1\over 2}
\mbox{Tr}(R_\mu R^\mu)-{1\over 16}\mbox{Tr}
([R_\mu,R_\nu][R^\mu,R^\nu]),
\label{lag}
\ee
where we have used scaled units of energy and length, and the baryon density ${\cal B}$, whose spatial integral gives the
 integer-valued
 baryon number $B$, is given by
\be
{\cal B}=-\displaystyle{1\over 24\pi^2}\epsilon_{ijk}\mbox{Tr}(R_iR_jR_k)\,.
\ee
As throughout this letter, we take latin indices to run over the
spatial values $1,2,3$.
The corresponding Faddeev-Bogomolny bound on the 
energy $E$ is simply $E\geq |B|\times 12\pi^2$.


\section{An ansatz with hexagonal symmetry}
\news

Following a suggestion of Atiyah, it has recently been proved by Jarvis \cite{Ja}
that the space of SU(2) BPS monopoles is diffeomorphic to the space of equivalence
classes of rational maps between Riemann spheres. These rational maps
arise from the monopole as the scattering data of a linear operator when considered
along all possible lines emanating from a chosen origin.
Through the use of a new ansatz for Skyrme fields \cite{HMS}, it has been possible
to use Jarvis rational maps to construct good approximations to the known minimum
energy Skyrmions. The fact that Jarvis maps, being the scattering
data along radial lines, are relevant reflects the feature
that these Skyrmions are shell-like structures.

In this letter we are concerned with a two-dimensional lattice of Skyrmions, which is 
clearly not a shell-like configuration, making Jarvis maps inappropriate for this
application. However, an older diffeomorphism of Donaldson \cite{Do}, between SU(2)
monopoles and based rational maps, is appropriate. These Donaldson maps arise as the
scattering data of a linear operator considered along all lines in a chosen direction.
This definition requires a decomposition of $\R^3$ into $\R\times\C$, which is
exactly the situation for a lattice configuration, where the complex plane refers
to the plane of the lattice and the real coordinate is the height above the lattice.
To be precise, since we wish to consider an infinite lattice, the Donaldson map will
not be rational, but instead is required to be merely meromorphic; which may be
regarded as an infinite limit of a rational map.

A consideration of the ansatz introduced in ref.\cite{HMS}, together with the
modifications discussed above, leads us to the following Skyrme field ansatz
\be
U(x_1,x_2,x_3)=\exp\bigg(\frac{if}{1+\vert W\vert^2}(W\tau_-+{\bar W}\tau_+
+(1-\vert W\vert^2)\tau_3)\bigg)\,,
\label{ansatz}
\ee
where $\tau_i$ denote the Pauli matrices with $\tau_\pm=\tau_1\pm i\tau_2$,
$W\in$\CP$^1$ is a holomorphic function of $z=x_1+ix_2$, and
$f\in\R$ is a function of $x_3.$

The lattice occupies the $x_1x_2$ plane, in which we use the complex coordinate $z.$
Thus from the above ansatz we see that the direction of the vector of pion fields
is determined by the \CP$^1$ field $W$, given the position in the lattice, whereas
the length of the vector of pion fields is determined by the profile function $f$,
given the height above the lattice.

The next issue to address is that of the boundary conditions on the Skyrme field $U$
and hence on the functions $f$ and $W.$ To have a periodic lattice there must exist
complex constants $\Omega_1$ and $\Omega_2$ which are the fundamental periods of
the lattice, that is,
\be
U(z+n\Omega_1+m\Omega_2,x_3)=U(z,x_3) \ \ \ \ \forall \ n,m\in\Z\,.
\ee
Let $T^2$ denote the associated fundamental parallelogram, that is,
the torus given by the region in the complex plane with vertices
$0,\Omega_1,\Omega_2,\Omega_1+\Omega_2$ and opposite edges identified.
We can now restrict our analysis to the region of $\R^3$ given by $\R\times T^2$,
with the field in the remaining regions determined by periodicity.
Thus we see from our ansatz (\ref{ansatz}) that $W$ is required to be a holomorphic
map $W: T^2\mapsto \CP^1.$

To consider the boundary conditions in the direction orthogonal to the plane
of the lattice we need to recall our motivation.
The lattice is being thought of as an infinite limit of the shell-like Skyrmions
containing pentagons and hexagons. Thus, in approaching this limit, we imagine the lattice as being a part of the bottom of a larger and larger shell and 
hence below the lattice is the outside of the shell, where the Skyrme field
tends to the vacuum, giving the boundary condition
$\lim_{x_3\rightarrow-\infty}U=\identity_2.$
However, above the lattice is the inside of the shell, where the Skyrme field
is approaching the negative vacuum associated with the centre of the Skyrmion,
so the appropriate boundary condition is 
$\lim_{x_3\rightarrow+\infty}U=-\identity_2.$
Note that this implies that our Skyrme lattice is a domain wall, separating regions
of differing vacuum values.
Examination of our ansatz (\ref{ansatz}) now reveals that the boundary conditions
for the profile function $f(x_3)$ read
\be
f(-\infty)=0\,, \ f(\infty)=\pi\,.
\label{bc}
\ee

To compute the baryon number and energy of the Skyrme field (\ref{ansatz})
in a fundamental section we shall follow the approach of ref.\cite{HMS}.
The strain tensor, defined as
\be
D_{ij}=-{1 \over 2}\mbox{Tr}(R_iR_j)\,,
\ee
is symmetric and positive semi-definite. If it has eigenvalues $\lambda_1^2,
\lambda_2^2, \lambda_3^2$ then the Skyrme energy density ${\cal E}$
and baryon density ${\cal B}$ are given by
\bea
&{\cal E}&=\lambda_1^2 +\lambda_2^2 +\lambda_3^2+ \lambda_1^2\lambda_2^2+ \lambda_2^2\lambda_3^2+
\lambda_1^2\lambda_3^2 \,,\\
&{\cal B}&=\lambda_1\lambda_2\lambda_3 / 2\pi^2.
\eea
For the ansatz (\ref{ansatz}) the strain in the direction normal to the lattice
is orthogonal to the two strains in the directions of the lattice, which are equal.
Therefore the $\lambda_i$ may be interpreted as the strains in the $x_i$ directions, making
it is easy to show that
\be
\lambda_3=f'\,, \ \ \lambda_1=\lambda_2=2J\sin f\,,
\label{strains}
\ee
where we have defined the quantity
\be
J=\frac{\vert \partial_z W\vert}{1+\vert W\vert^2}\,.
\ee
Substituting the expressions for the strains (\ref{strains}) into those for
the energy and baryon density, we arrive at the result
\bea
&{\cal E}&=f'^2+8J^2(f'^2+\sin^2f)+16J^4\sin^4f \,,\label{eden}\\ 
&{\cal B}&=\frac{2}{\pi^2}J^2f'\sin^2f\,.\label{bden}
\eea

We now wish to use these densities to compute the energy $E$ and
baryon number $B$ in a fundamental section of the lattice, by
integrating over the region $x_3\in(-\infty,\infty)$ and $(x_1,x_2)\in T^2.$
To do this we note that since $W$ is a map $W: T^2\mapsto \CP^1,$ then it
has an associated integer, $k$, which is its degree. Explicitly, $k$ is given
by integrating over the torus the pullback under $W$ of the Fubini-Study area
2-form on \CP$^1$, which in this case gives
\be
k=\frac{1}{\pi}\int_{T^2} J^2\ dx_1dx_2\,,
\label{deg}
\ee
since $W$ is a holomorphic function of $z.$

Using (\ref{bden}) it is now easy to see that the baryon number is equal
to the degree $k$, since
\be
B=\frac{2}{\pi^2}\int_{-\infty}^{\infty} f'\sin^2f\ dx_3\int_{T^2} J^2\ dx_1dx_2
=\frac{k}{\pi}\bigg[f-\frac{1}{2}\sin 2f\bigg]_{-\infty}^{\infty}=k\,,
\ee
where we have used the expression (\ref{deg}) and the boundary conditions (\ref{bc}).

In calculating the energy it will be useful to introduce a scale parameter $\mu$
by writing $u=x_3/\mu$ and setting $f(x_3)=g(u).$
Then, if $A$ is the area of the fundamental torus $T^2$, integrating the density
(\ref{eden}) gives
\be
E=\int_{-\infty}^{\infty}dx_3\int_{T^2} dx_1dx_2\ {\cal E}
=\frac{A}{\mu}E_1+\frac{1}{\mu}E_2+\mu E_3 +\frac{\mu}{A}E_4
\label{energy}
\ee
where the $E_i$'s are the following integrals over $u$
\bea
&E_1&=\int_{-\infty}^{\infty} g'^2\ du\,, \nonumber\\
&E_2&=8\pi k\int_{-\infty}^{\infty} g'^2\sin^2g\ du\,, \label{es}\\
&E_3&=8\pi k\int_{-\infty}^{\infty} \sin^2g\ du\,, \nonumber \\
&E_4&=16 {\cal I}\int_{-\infty}^{\infty} \sin^4g\ du\,. \nonumber
\eea
The only remaining dependence on the map $W$ is the quantity ${\cal I}$, which
is defined as
\be
{\cal I}=A\int_{T^2} J^4\ dx_1dx_2\,,
\label{defi}
\ee
and has the important property that it is independent of $A.$

The scale $\mu$ and area $A$ can now be determined, in terms of the $E_i$'s,
by minimization of the energy (\ref{energy}). Requiring $\frac{\partial E}{\partial \mu}
=\frac{\partial E}{\partial A}=0,$ gives the result
\be
\mu=\sqrt{E_2/E_3}\,, \ \ \ A=\sqrt{E_2E_4/E_1E_3}\,,
\label{muanda}
\ee
and hence a minimized energy of
\be
E=2(\sqrt{E_1E_4}+\sqrt{E_2E_3})\,.
\label{minenergy}
\ee

To proceed further we now need an explicit expression for the map
$W(z).$ To obtain a holomorphic map from the torus, we
take $W$ to be an elliptic function of $z.$ Exactly which elliptic function 
to take is 
determined by the fact that we wish to construct a hexagonal lattice, so
we require a fundamental period parallelogram which has a $60^\circ$ angle
between the two fundamental periods. The appropriate elliptic function is
the Weierstrass function $\wp(z)$ satisfying
\be
\wp'^2=4(\wp^3-1)\,,
\ee
which has real period $\Omega_1=\Gamma(\frac{1}{6})\Gamma(\frac{1}{3})/(2\sqrt{3\pi})$
and imaginary period given by $\Omega_2=\Omega_1(1+i\sqrt{3})/2.$ From this we see that
the period lattice is equilaterally triangular, and thus we have the desired 
$60^\circ$ angle. Obviously we can scale both the elliptic function and its
argument and still retain the above desired property, hence we take
\be
W(z)=c\wp(z/\alpha)\,,
\label{dag}
\ee
where $c$ and $\alpha$ are arbitrary real constants. Note that by the inclusion
of the factor $\alpha$ we must now rescale the fundamental torus. Furthermore,
for later computational purposes it is convenient to work with a rectangular
fundamental torus, which is achieved by taking it to be the $T^2$ given by
$(x_1,x_2)\in [0,\alpha\Omega_1]\times[0,\alpha\sqrt{3}\Omega_1].$
As this torus contains two fundamental parallelograms and the $\wp$-function has
a double pole in each of these then, by counting preimages, we see that the
degree of the map in this case is $k=4.$

For a given $k$, the energy $E$ is minimized by minimizing the value of the
integral ${\cal I}$ defined in (\ref{defi}). This is clear since its only
appearance is as a coefficient in front of the positive term $E_4$ in (\ref{es}).
Recall that ${\cal I}$ is independent of the area $A$, and hence $\alpha$ since the two
are simply related by
\be
A=\sqrt{3}\alpha^2\Omega_1^2\,.
\ee
Computing ${\cal I}$ for the one-parameter family of maps (\ref{dag}), given
by varying $c$ for any fixed $\alpha$, we find that its minimum value
is ${\cal I}\approx 193$, which is attained when $c\approx 0.7.$

In order to continue with an analytical treatment we now make an ansatz
for the profile function $g(u)$, which we shall see turns out to be a 
reasonably good choice. We choose the sine-Gordon kink profile function
\be
g(u)=2\mbox{arctan} \ e^u\,,
\label{kink}
\ee
which has the advantage that all the integrals in (\ref{es}) can be performed exactly.
The results are
\be
E_1=2\,, \ E_2=128\pi/3\,, \ E_3=64 \pi\,, \ E_4=64{\cal I}/3\,,
\ee
from which we find that the scale and area are
\be
\mu=\sqrt{2/3}\,, \  \ \ A=\frac{8}{3}\sqrt{{\cal I}}\,,
\ee
and using (\ref{minenergy}) the energy is
\be
E=16\sqrt{2/3}(\sqrt{{\cal I}}+8\pi)\,.
\ee
Recalling that $B=k=4$ we thus compute that the energy per baryon is
\be
E/B=1.076\times 12\pi^2.
\label{epblat}
\ee

In discussing the energy of various configurations it will be useful to
define $\Delta$ to be the percentage excess over the Faddeev-Bogomolny bound ie.
\be
\Delta=\bigg(\frac{E-12\pi^2\vert B\vert}{12\pi^2\vert B\vert}\bigg)\times 100\% .
\label{excess}
\ee
Thus, our ansatz for the lattice has $\Delta_{\widetilde {\rm lat}}=7.6\%.$

\section{Skyrmion Architecture}
\news

In this section we discuss the accurately computed 
 energies of various configurations,
using a numerical relaxation of the full nonlinear Skyrme model.
 Using this information we speculate on the kinds of structures 
which may form for Skyrmions of high charge.

In the previous section we were able to compute an explicit Skyrme field
which describes a lattice with a relatively small excess energy. However,
some approximations were made, for example, the choice of the profile function
(\ref{kink}) and so the energy of the true lattice will be lower than the
ansatz value of $\Delta_{\widetilde {\rm lat}}=7.6\%.$ To determine the true value, we take
the ansatz field as a starting configuration in a numerical relaxation
computation using the code described in detail in ref. \cite{BSc}. The 
simulations were performed on a grid containing 
$100\times 100\times 58$ points and periodic
boundary conditions imposed in appropriate directions. The computation
of the initial excess confirmed the value to be $\Delta_{\widetilde {\rm lat}}=7.6\%$
but after relaxation, which also involved some minor rescaling to be
sure to obtain the minimum energy lattice, the excess of the true lattice
was found to settle down to $\Delta_{\rm lat}=6.1\%.$

In Fig.1 we display a surface of constant baryon density for the hexagonal
lattice. The hexagonal structure is clearly visible, with the baryon density
isosurface having a hole in the centre of each of the hexagonal faces.
Note that the section of lattice we are considering contains exactly eight full hexagons
and is of baryon number four, computed to be $B=3.84$ on the discretized grid. The fact that each hexagon may be thought of
as having baryon number one-half is the expected infinite limit of the
polyhedron structure discussed in the introduction, where a charge $B$
Skyrmion is found to have $2(B-1)$ faces.

As mentioned in the introduction, the excess energy of the lattice is
lower than that found for any finite charge Skyrmion. For a single
Skyrmion the excess is $\Delta_{1}=23\%$, whereas for the configurations
up to charge nine the minimum excess occurs for $B=9$ which has 
$\Delta_{9}=9.8\%$
\cite{BSb}. A relaxation of the charge seventeen buckyball configuration
\cite{HMS} gives the smallest known value $\Delta_{17}=7.2\%$ \cite{BSc}.
Thus our results are all consistent with the fullerene picture of Skyrmions
where, at least for Skyrmions of modest charge, the structure is a shell of
pentagons and hexagons.

The fact that $\Delta_{\rm lat}$ is so low is encouraging for the possibility
of Skyrmion shells at even higher charge, however it is not low enough
to conclude that shell-like structures continue indefinitely. This is 
because of the existence of the Skyrme crystal \cite{KS,CJJVJ}, which is
a configuration that is periodic in all three space dimensions and consists
of a crystal of half-Skyrmions. A Fourier series analysis \cite{KS} approximates
the energy excess of the crystal to be $\Delta_{\widetilde {\rm cry}}=3.8\%$ and
a simple analytical formula exists which gives a good approximation to 
the fields of the Skyrme crystal \cite{CJJVJ}. Using this as an initial
configuration in our relaxation scheme, we find a true energy for the
Skyrme crystal of  $\Delta_{\rm cry}=3.6\%.$ A surface of constant baryon
density is shown in Fig.2. The half-Skyrmion structure of the crystal
is evident and it is clear that this configuration is of a very
different type to that of the lattice shown in Fig.1.

Naively, the fact that  $\Delta_{{\rm cry}}<\Delta_{\rm lat}$ suggests that there could
be some value of the Skyrmion charge, $B_{\rm crit}$, at which the shell-like
structures will be replaced by more three-dimensional configurations,
but the details of how and when this might take place are unclear.
However, a simple comparison of $\Delta_{{\rm cry}}$ and $\Delta_{\rm lat}$ measures the volume effect, but since these two values are not so different
the crucial factor will be an area effect, which is associated with
the fact that a finite portion of the Skyrme crystal needs to
be smoothed off at the edges. One suitable candidate for completing
the edge of the Skyrme crystal is a configuration like the face
of the $B=4$ cubic Skyrmion. However, since this has an excess
of $\Delta_{4}=12\%$, quite a substantial piece of the crystal
needs to be included before such a high penalty for its boundary
could be accommodated. In contrast, a large shell structure
only ever requires twelve pentagon defects and they can be included
at a relatively small cost, as demonstrated by the $B=17$ buckyball
with $\Delta_{17}=7.2\%.$ All this suggests that even if the Skyrme
crystal structure appears at some charge $B_{\rm crit}$ then this may well
be very high. Therefore, there is at least the possibility of a range
of modest charges where other fullerene structures may exist.
Other exotic possibilities include analogues of the bucky-tubes 
(long thin configurations comprised of spirals of hexagons with
caps containing pentagons) and also shells inside shells.

If the above structure change to the Skyrme crystal does
indeed occur then for $B\ge B_{\rm crit}$ then
the fullerene analogy will be lost, since the Skyrme crystal is
a configuration with valency six. However, it is possible that the
known Skyrme crystal is not the minimum energy crystal structure.
If, for example, a configuration exists with the structure of the
diamond lattice, then the similarities with carbon atoms could be
maintained. This is a four-valent lattice with tetrahedral symmetry,
but would not be as simple as the tetrahedral lattice formed from
individual Skyrmions in the nearest neighbour attractive channel, 
which relaxes to the cubically symmetric Skyrme crystal.
However, it is a difficult task to investigate this possibility
since a Skyrme field with the correct properties needs to be
found before numerical simulations can be performed. Using the
ideas from rational maps we were able to create a configuration
to study the lattice and perhaps a similar technique could be employed
to study other crystals.

\section{Conclusions}
\news 

We have introduced an ansatz for a Skyrme lattice with hexagonal symmetry
and used it as an initial condition in a numerical relaxation
of the full nonlinear equations of the Skyrme model.
The result is further evidence to support the analogy between Skyrmions
and fullerene shells in carbon chemistry.

Further work is required to address the issue of the structure of high
charge Skyrmions. One approach is to collide Skyrmions, but it may also
prove useful to use the Skyrme lattice, or at least a part of it.
The interaction of two such lattices may shed some light on the
subject of shell formation and it would also be interesting to
study the scattering of Skyrmions off the lattice. These issues are
currently under investigation.

Given the connection between Skyrmions, monopoles and rational maps \cite{HMS}
it seems likely that a similar lattice of monopoles will exist. 
In this letter we chose a specific elliptic function in our ansatz in order
to obtain a hexagonal lattice. Other Skyrme fields can be obtained which
correspond to different lattices by choosing other elliptic functions,
though we expect them to have higher energy per baryon than the one considered
here \footnote{We have, in fact, also repeated the same procedure for a square lattice, 
which is quad-valent. The calculated excess $\Delta_{\rm sq}=6.8\%$, larger than that
 for the trivalent, hexagonal lattice, illustrating that the trivalent is more
 energetically favourable.}. However, in the monopole context all types of lattices
 would be on
an equal footing since monopoles are BPS solitons. It would therefore be 
interesting to see if the Donaldson correspondence could be extended to 
infinite charge monopole lattices, with each characterized by an elliptic function.

\section*{Acknowledgements}
\news

We thank Conor Houghton, Chris Barnes, Kim Baskerville, Nick Manton and Neil Turok for
 useful discussions. RAB acknowledges the support of PPARC Postdoctoral fellowship
 grant GR/K94799 and PMS thanks the Nuffield Foundation for a newly appointed 
lecturer award. We also acknowledge the use of the SGI Origin 2000 and Power
 Challenge at DAMTP in Cambridge supported by the HEFCE, SGI, PPARC, the Cambridge
 Relativity rolling grant and EPSRC Applied Mathematics Initiative grant GR/K50641. 

\def\jnl#1#2#3#4#5#6{\hang{#1 [#2], {\it #4\/} {\bf #5}, #6.}}
\def\jnltwo#1#2#3#4#5#6#7#8{\hang{#1 [#2], {\it #4\/} {\bf #5}, #6;{\bf #7} #8.}}
\def\prep#1#2#3#4{\hang{#1 [#2],`#3', #4.}} 
\def\proc#1#2#3#4#5#6{{#1 [#2], in {\it #4\/}, #5, eds.\ (#6).}}
\def\book#1#2#3#4{\hang{#1 [#2], {\it #3\/} (#4).}}
\def\jnlerr#1#2#3#4#5#6#7#8{\hang{#1 [#2], {\it #4\/} {\bf #5}, #6.
{Erratum:} {\it #4\/} {\bf #7}, #8.}}
\def\prl{Phys.\ Rev.\ Lett.}
\def\pr{Phys.\ Rev.}
\def\pl{Phys.\ Lett.}
\def\np{Nucl.\ Phys.}
\def\prp{Phys.\ Rep.}
\def\rmp{Rev.\ Mod.\ Phys.}
\def\cmp{Comm.\ Math.\ Phys.}
\def\mpl{Mod.\ Phys.\ Lett.}
\def\apj{Ap.\ J.}
\def\apjl{Ap.\ J.\ Lett.}
\def\aap{Astron.\ Ap.}
\def\cqg{Class.\ Quant.\ Grav.} 
\def\grg{Gen.\ Rel.\ Grav.}
\def\mn{M.$\,$N.$\,$R.$\,$A.$\,$S.}
\def\ptp{Prog.\ Theor.\ Phys.}
\def\jetp{Sov.\ Phys.\ JETP}
\def\jetpl{JETP Lett.}
\def\jmp{J.\ Math.\ Phys.}
\def\zpc{Z.\ Phys.\ C}
\def\cupress{Cambridge University Press}
\def\oup{Oxford University Press}
\def\pup{Princeton University Press}
\def\wss{World Scientific, Singapore}

\section*{Figure Captions}  

\noindent Fig.~1. Baryon density isosurface for the Skyrme lattice. 
The section displayed has baryon number four and contains effectively 
eight hexagons
\bigskip

\noindent Fig.~2. Baryon density isosurface for the Skyrme crystal.
 Each corner contains a half-skyrmion and the total baryon number shown 
is four. If the threshold for the isosurface were to be 
increased the corners would be connected by links of lower baryon 
density into a crystalline structure.

\bigskip


\begin{figure}[ht]
\begin{center}
\leavevmode
{\epsfxsize=16cm \epsffile{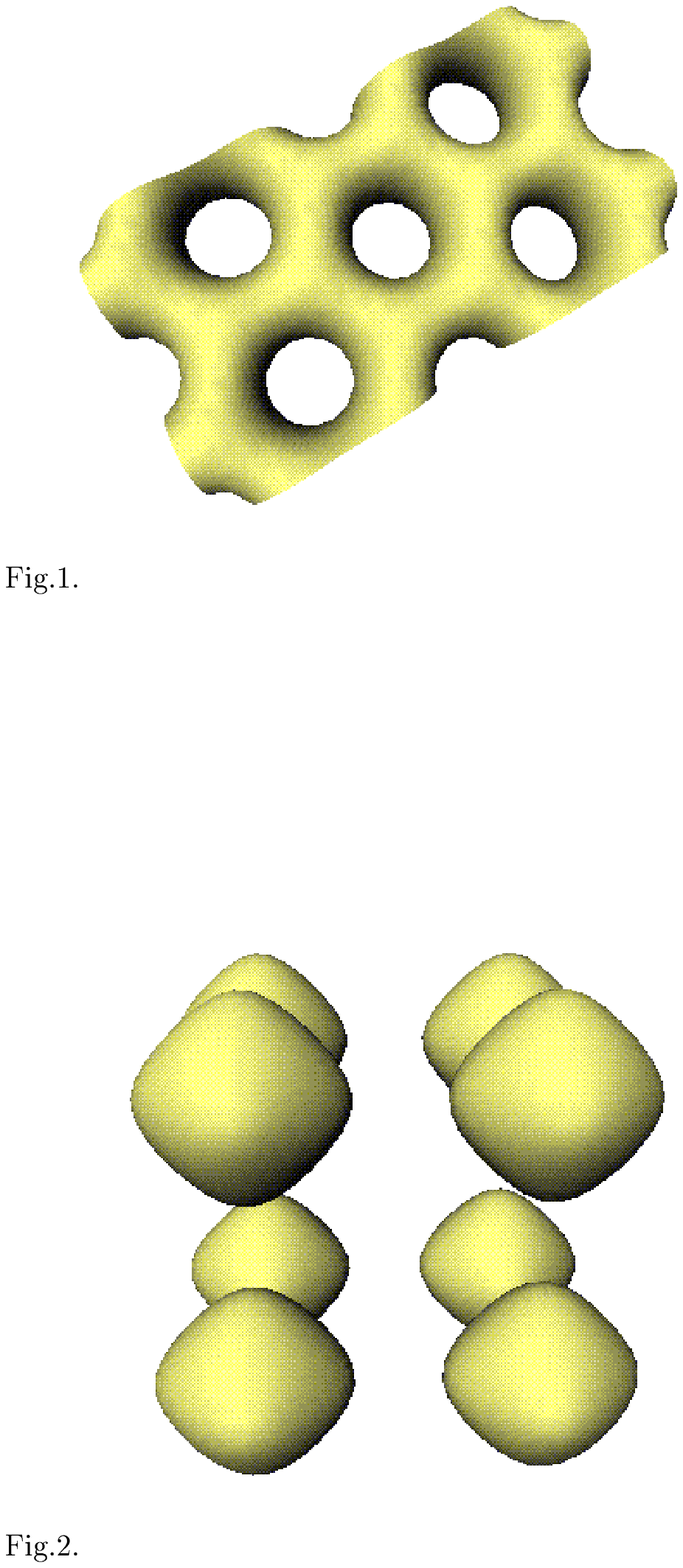}}
\end{center}
\end{figure}

\end{document}